\documentclass[a4paper,11pt]{article}
\pdfoutput=1 
\usepackage{graphicx} 

\usepackage{jinstpub} 

\usepackage{lineno}

\usepackage{algorithmic}
\usepackage[acronym]{glossaries}
\glsdisablehyper 

\usepackage{subcaption}
\usepackage{microtype}
\newacronym{lhc}{LHC}{Large Hadron Collider}
\newacronym{hl-lhc}{HL-LHC}{High-Luminosity Large Hadron Collider}
\newacronym{itk}{ITk}{Inner Tracker}
\newacronym{dcs}{DCS}{Detector Control System}
\newacronym{mops}{MOPS}{Monitoring of Pixel System}
\newacronym{mhfb}{MH Mock-up}{MOPS-Hub Mock-up}
\newacronym{mh}{MH}{MOPS-Hub}
\newacronym{fpga}{FPGA}{Field-Programmable Gate Array}
\newacronym{can}{CAN}{Controller Area Network}
\newacronym{cic}{CIC}{CAN Interface Card}
\newacronym{asic}{ASIC}{Application-Specific Integrated Circuit}
\newacronym{jcop}{JCOP}{Joint COntrols Project}
\newacronym{opcua}{OPC UA}{Open Platform Communications Unified Architecture}
\newacronym{daq}{DAQ}{Data Acquisition}
\newacronym{gbt}{GBT}{GigaBit Transceiver}
\newacronym{io}{I/O}{Input/Output}
\newacronym{sram}{SRAM}{Static Random-Access Memory}
\newacronym{usb}{USB}{Universal Serial Bus}
\newacronym{uart}{UART}{Universal Asynchronous Receiver-Transmitter}
\newacronym{pp3}{PP3}{Patch Panel 3}
\newacronym{pp0}{PP0}{Patch Panel 0}
\newacronym{atlas}{ATLAS}{A Toroidal LHC Apparatus}
\newacronym{alice}{ALICE}{A Large Ion Collider Experiment}
\newacronym{cms}{CMS}{Compact Muon Solenoid}
\newacronym{lhcb}{LHCb}{Large Hadron Collider beauty}
\newacronym{cern}{CERN}{European Council for Nuclear Research}
\newacronym{pcb}{PCB}{Printed Circuit Board}
\newacronym{irb}{IRB}{Intermediate Readout Board}
\newacronym{emci}{EMCI}{Embedded Monitoring and Control Interface}
\newacronym{emp}{EMP}{Embedded Monitoring Processor}
\newacronym{pdo}{PDO}{Process Data Object}
\newacronym{sdo}{SDO}{Service Data Object}
\newacronym{qc}{QC}{Quality Control}
\newacronym{mdts}{MDTs}{Drift Tubes}

\title{\boldmath A Comparative Analysis of the \\ CERN ATLAS ITk MOPS Readout: \\ A Feasibility Study on Production and Development Setups}

\author[a,1]{Lukas Flad,\note{Corresponding author.}}
\author[a]{Felix Sebastian Nitz}
\author[a]{and Tobias Krawutschke}

\affiliation[a]{Faculty of Information, Media and Electrical Engineering,\\University of Sciences Cologne,\\Germany}

\emailAdd{lukas.flad@th-koeln.de}

\abstract{
The upcoming High-Luminosity upgrade of the Large Hadron Collider (LHC) necessitates a complete replacement of the ATLAS Inner Detector with the new Inner Tracker (ITk). This upgrade imposes stringent requirements on the associated Detector Control System (DCS), which is responsible for the monitoring, control, and safety of the detector. A critical component of the ITk DCS is the Monitoring of Pixel System (MOPS), which supervises the local voltages and temperatures of the new pixel detector modules. This paper introduces a dedicated testbed and verification methodology for the MOPS readout, defining a structured set of test cases for two DCS-readout architectures: a preliminary Raspberry~Pi–based controller, the “MOPS-Hub Mock-up” (MH Mock-up), and the final production FPGA-based “MOPS-Hub” (MH). The methodology specifies the measurement chain for end-to-end latency, jitter, and data integrity across CAN and UART interfaces, including a unified time-stamping scheme, non-intrusive signal taps, and a consistent data-logging and analysis pipeline. This work details the load profiles and scalability scenarios (baseline operation, full-crate stress, and CAN Interface Card channel isolation), together with acceptance criteria and considerations for measurement uncertainty to ensure reproducibility. The objective is to provide a clear, repeatable procedure to qualify the MH architecture for production and deployment in the ATLAS ITk DCS. A companion paper will present the experimental results and the comparative analysis obtained using this testbed.
}


\arxivnumber{XXXX.XXXX} 

\collaboration{on behalf of the ATLAS ITk Collaboration}

\begin{document}
\maketitle




\flushbottom
\section{Introduction}
The ATLAS experiment \cite{aadATLASExperimentCERN2008} at the \gls{lhc} is undergoing a major upgrade for the \gls{hl-lhc} era \cite{aberleHighLuminosityLargeHadron2020}. The \gls{hl-lhc} will deliver a significantly higher integrated luminosity, leading to a much harsher radiation environment and higher data rates. To cope with these new conditions, the current ATLAS Inner Detector will be replaced by an all-silicon \gls{itk} \cite{TechnicalDesignReport2017a, cern.geneva.thelhcexperimentscommitteeTechnicalDesignReport2017a, thompsonATLASITkPixel2025, mobiusPixelDetectorStudies2023, chubinidzeITkPixelSystem2025}.

The \gls{dcs} is an essential supervision of the experiment, ensuring the safe and reliable operation of the detector components \cite{poyDetectorControlSystem2008}. For the \gls{itk}, the \gls{dcs} must manage thousands of power supplies, monitor environmental parameters, and handle potential fault conditions in real-time. The \gls{mops} is a key sub-system of the \gls{itk} \gls{dcs}, tasked with reading out the local voltages and temperatures from the pixel modules via a custom radiation-tolerant \gls{asic} \cite{ahmadMonitoringPixelSystem2022, ahmadSecondGenerationMonitoring2023b}. The \gls{mops} \glspl{asic} transmit this data via a specially designed 1.2V physical layer \gls{can} bus, used due to radiation hardness demands. The detector upgrade requires MOS transistors of thinner gate-oxide, than the standard 5V \gls{can} bus and its MOS transistors of thicker gate-oxides allow \cite{ahmadDescriptionManualUsea}.

This data must be reliably read out from the ATLAS cavern, aggregated, and made available to the central \gls{dcs} infrastructure. The development of the readout system has followed a staged approach. Initially a first setup, called \gls{mhfb} (internally known as MOPS-Hub for beginners) crate was developed, which uses a Raspberry Pi as its main processing unit \cite{lezkiMOPSHubBeginnersDocumentationa}. This system was intended as an intermediate solution to facilitate early hardware and software validation. The final proposed architecture, the \gls{mh} crate, utilizes a robust, custom-built platform featuring a powerful \gls{fpga} as its controller \cite{qameshFPGAbasedDataAggregator2024, qameshSystemIntegrationATLAS2023c}. Before proceeding with large-scale production, a rigorous, quantitative validation methodology is required to qualify the \gls{mh} architecture.

It is important to note that beyond the communication performance metrics evaluated in this work, the final production hardware must operate reliably within the radiation environment of the ATLAS cavern. While a detailed analysis of radiation hardness is outside the scope of this comparative study, the design of the MH accounts for these effects. As detailed in dedicated irradiation studies, a safety factor of 3 is applied to these simulated levels during hardware qualification to ensure long-term operational stability~\cite{qameshFPGAbasedDataAggregator2024}. There, the methodology focuses specifically on the validation of the data readout chain's timing performance and integrity, which is a complementary and equally critical step for final system approval.

This paper presents such a methodology, centered on a dedicated testbed designed to precisely evaluate and compare the \gls{mhfb} and \gls{mh} systems. The key contributions of this work are:
\begin{itemize}
    \item A single-clock microcontroller-based testbed for measuring sub-millisecond latency and jitter across the 1.2\,V \gls{can} to \gls{uart} data path, overcoming the limitations of standard PC-based tools.
    \item A comprehensive three-stage test plan defining procedures for baseline performance, full-crate stress testing, and \gls{cic} channel isolation verification.
    \item A complete data-logging pipeline, including hardware signal taps, firmware for data capture, and Python scripts for automated analysis and visualization.
\end{itemize}

The remainder of this paper is structured as follows: Section \ref{sec:related_work} reviews related \gls{dcs} and readout technologies. Section \ref{sec:methodology} describes the architecture of the \gls{mops} readout chain and the testbed design. Section \ref{sec:validity} discusses the measurement validity. Section \ref{sec:evaluation} outlines the detailed test procedures and quantitative metrics. Section \ref{sec:conclusion} concludes the paper and Section \ref{sec:future_work} outlines future work.

\section{Related Work}
\label{sec:related_work}
The development of \glspl{dcs} for large-scale high-energy physics experiments is a complex challenge, balancing requirements for reliability, scalability, and radiation tolerance. The control systems for the \gls{lhc} experiments, including ATLAS, CMS, ALICE and LHCb, are typically large, distributed systems based on commercial and custom-built hardware and software \cite{swobodaDetectorControlSystem1999, chochulaChallengesALICEDector2018, alessioLHCbDataAcquisition2014, barillereLHCGasControl2007, adamComputingShiftsMonitor2017, chatrchyanCMSExperimentCERN2008}. A common framework, the \gls{jcop}, was established to unify the control systems architecture across the \gls{lhc} experiments, promoting the use of standard components like \gls{opcua} for data exchange \cite{holmeJCOPFramework2005, eckerEMCIEMPDevelopmentsExperience2025}.

The validation strategies for these complex detector systems often focus on holistic commissioning using cosmic rays and single-beam data, as demonstrated by the extensive campaigns for the ATLAS Muon Spectrometer and the Tile Calorimeter~\cite{ATLASCollaboration2010_Muon, ATLASCollaboration2010_TileCal}. These studies validated overall system performance, including trigger efficiency, alignment, and energy scale calibration, confirming readiness for LHC collisions. While these approaches are essential for final system integration, they typically present the performance of the final hardware without detailing a quantitative, comparative methodology for evaluating prototype versus production architectures on specific communication metrics like sub-millisecond latency and jitter.

Furthermore, general testing methodologies in high-energy physics range from component-level quality assurance to system-level integration tests. For instance, the mass production of Monitored \gls{mdts} for the Muon Spectrometer involved rigorous, automated tests on each individual component, measuring parameters such as gas leak rates, wire tension, and dark current to ensure they met strict design specifications~\cite{Aprile2002_MDT}. This work bridges the gap between such component-level validation and final system commissioning by presenting a dedicated, repeatable testbed designed to explicitly quantify the communication performance and reliability of the data readout chain—a critical intermediate step not extensively detailed in broader commissioning papers.

Furthermore, general testing methodologies in high-energy physics range from component-level quality assurance to system-level integration tests. For instance, the mass production of Monitored \gls{mdts} for the Muon Spectrometer involved rigorous, automated tests on each individual component, measuring parameters such as gas leak rates, wire tension, and dark current to ensure they met strict design specifications~\cite{Aprile2002_MDT}. The methodology presented in this paper serves a dual purpose within this validation framework, addressing both intermediate component quality control and production system qualification. First, the testbed and the \gls{cic} isolation test (see Section \ref{subsec:case_3}) are integral to the \gls{qc} procedure for the \gls{cic} boards, verifying critical hardware isolation requirements before they are integrated into the final crates. Second, the baseline and stress tests (see Sections \ref{subsec:case_1} and \ref{subsec:case_2}) are designed to provide the definitive performance validation of the final FPGA-based MOPS-Hub architecture, qualifying its communication robustness and scalability before large-scale production for the Run 4 upgrade. Thus, this work defines a crucial intermediate validation stage between individual component checks and full system commissioning, focusing specifically on quantifying the communication performance and reliability of the data readout chain.

The use of \glspl{fpga} in \gls{daq} and \gls{dcs} applications is widespread due to their parallel processing capabilities and deterministic, low-latency performance \cite{almeRadiationTolerantSRAMFPGABased2008, dharaDesignDataAcquisition2012}. \glspl{fpga} are essential in trigger systems \cite{achenbachATLASLevel1Calorimeter2008, koulourisPhaseIIUpgradeATLAS2024} and front-end electronics where real-time processing is critical. In the context of \gls{dcs} \glspl{fpga} are often used for custom bus interfacing, data aggregation, and implementing fast feedback loops for equipment protection \cite{butkowskiFPGAFirmwareFramework2015b, frazierDemonstrationTimeMultiplexed2012}. The \gls{gbt} and Versatile Link projects at CERN provide radiation-hard, high-speed optical links for data transmission from the detector, often interfacing with \glspl{fpga} \cite{moreiraGBTProject2009, vaseyVersatileLinkCommon2012}. The custom hardware platform used in the \gls{mh} design is part of this ecosystem of custom electronics developed for the ATLAS upgrade.

\gls{can} bus is a robust serial communication protocol widely used in the automotive and industrial sectors \cite{ISO1189812015}. Its resilience to noise and built-in error detection mechanisms makes it suitable for control applications in harsh environments. At CERN, CANopen, a higher-level protocol based on \gls{can}, is used for \gls{dcs} applications \cite{nikielCANopenELMB2022}. For the ATLAS ITk upgrade, the \gls{mops} \gls{asic} employs a non-standard 1.2V low-voltage \gls{can} physical layer operating at 125 kbit/s. This design avoids the risks associated with the 5V MOS transistors of a standard \gls{can} bus physical layer architecture \cite{ahmadDescriptionManualUsea}, thereby making the system more resilient.

The use of single-board computers like the Raspberry Pi for prototyping has become popular due to their low cost and ease of use \cite{matheComprehensiveReviewApplications2024}. However, their application in critical, real-time systems is limited, as the default Linux-based operating system is not designed for the deterministic processing required. Performance analyses confirm that even with enhancements, the maximum warranted latencies are too high for critical tasks where timing faults are intolerable, making the platform suitable only for non-critical applications~\cite{carvalho2019raspberry}. This work directly addresses this limitation by establishing a methodology to quantitatively compare a Raspberry Pi-based prototype with a dedicated, \gls{fpga}-based system designed for high-reliability operation.

\section{Methodology and System Architecture}
\label{sec:methodology}

The core of this methodology is the precise measurement of the performance of two distinct data processing pipelines, \gls{mhfb} and \gls{mh}. The primary performance metric is the end-to-end latency ($\Delta t$) and data loss. Data losses are measured by the amount of data sent and what information can be found on the \gls{can} bus. For the latency, since the theoretical minimum response time is dictated by the 125~kbit/s \gls{can} bus, which is the slowest component in the link \cite{ahmadDescriptionManualUsea}, stringent performance benchmarks must be established. This includes a 7~ms (see Section \ref{subsec:case_1}) round-trip time for baseline operations, which ensures a high safety margin for the production system. Achieving this requires a detailed understanding of the system architecture and a carefully designed testbed to capture the necessary data without introducing significant measurement artifacts.

\subsection{MOPS Readout Chain}
\begin{figure}[htbp]
    \centering
    \begin{subfigure}{\columnwidth}
        \centering
        \includegraphics[width=\linewidth]{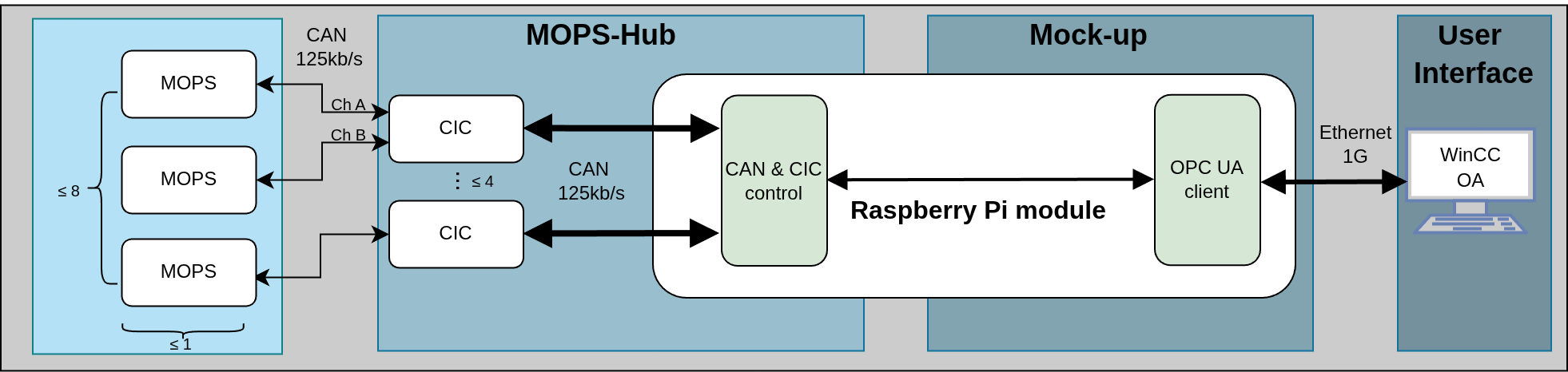}
        \caption{MOPS-Hub Mock-up Architecture}
        \label{fig:mhfb_arch}
    \end{subfigure}
    \vfill
    \begin{subfigure}{\columnwidth}
        \centering
        \includegraphics[width=\linewidth]{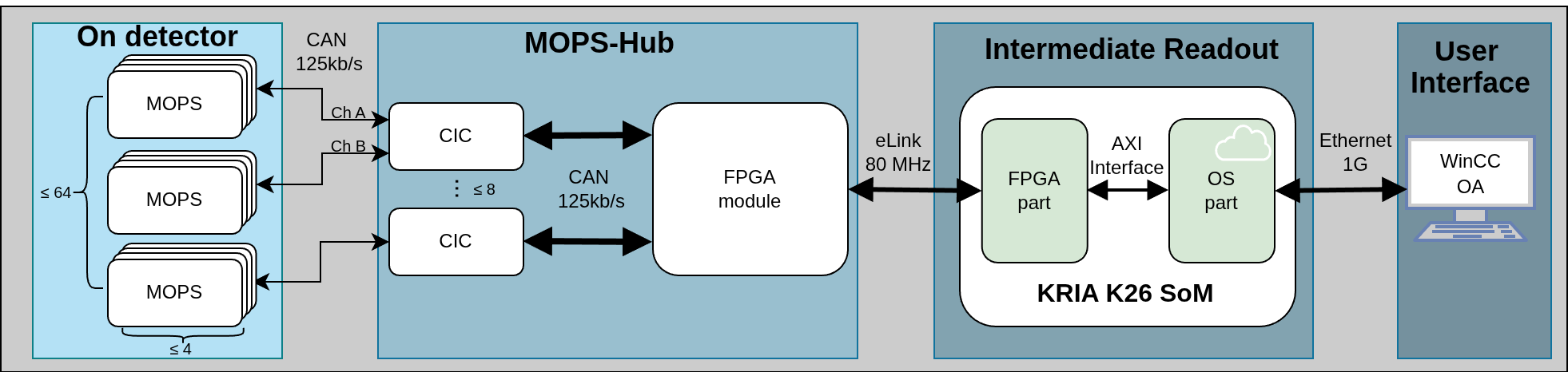}
        \caption{MOPS-Hub with Intermediate Readout Board (IRB) Architecture}
        \label{fig:mh_arch}
    \end{subfigure}
    \caption{Comparison of the MH Mock-up and MH on-detector test architectures.}
    \label{fig:system_comparison}
\end{figure}

The different data flows from the on-detector electronics to the control room, that were analyzed in this work, are illustrated in Figure \ref{fig:system_comparison}.

The main components of the readout chain are:
\begin{itemize}
    \item \textbf{\gls{mops} \gls{asic}}: An on-detector \gls{asic} that monitors the voltage and temperature of the detector for a group of pixel modules. The gathered data from the modules are transmitted over the \gls{can} buses. Up to four \gls{mops} \glspl{asic} can be on a single \gls{can} bus, identified by their unique NodeID (0-3) \cite{ahmadDescriptionManualUsea, qameshSystemIntegrationATLAS2023c}.
    \item \textbf{\gls{can} Bus}: A 1.2V differential \gls{can} bus operating at 125 kbit/s transmits data using the CANopen protocol from the \gls{mops} \glspl{asic} over distances up to 70 meter. This non-standard low-voltage physical layer was chosen specifically to avoid the risks associated with the 5V MOS transistors of a standard \gls{can} architecture, thereby making the system more resilient within the detector environment \cite{ahmadDescriptionManualUsea}.
    \item \textbf{\gls{cic}}: Located in the ATLAS cavern, the \gls{cic} provides galvanic isolation between the detector front-end and the readout electronics\cite{ATLASITkGrounding}. Each \gls{cic} has two independent channels (A and B).
    \item \textbf{Backplane}: A custom backplane distributes power and communication signals within a crate. It connects up to 8 \glspl{cic} to a central hub.
    \item \textbf{\gls{mh} / \gls{mhfb}}: The central component under test. It aggregates data from all \glspl{cic} and forwards it via e-Link \cite{bonaciniElink2009} (\gls{mh}) or Ethernet (\gls{mhfb}). The \gls{mhfb} uses a Raspberry Pi and can handle up to 4 \glspl{cic} (4 channels) \cite{Ecker2022MHFB}, while the \gls{mh} uses a custom \gls{fpga} \gls{pcb} designed for all 8 \glspl{cic} (16 channels).
    \item \textbf{Data Transmission}: The data path differs significantly between the two systems for the production testing phase. For both systems, a PC running AlmaLinux 9 with WinCC and an \gls{opcua} server is used for data visualization and control.
    In the \gls{mhfb} setup, the Raspberry Pi (client) sends aggregated data directly via Ethernet to the PC, which acts as an \gls{opcua} server. 
    In the \gls{mh} setup, the \gls{fpga} (client) sends data via a high-speed eLink to an \gls{irb}. The \gls{irb} then communicates with the same PC-based \gls{opcua} server. This setup is an intermediate step to validate the \gls{mh} crate before its integration into the final \gls{dcs} infrastructure, which will replace the \gls{irb} with dedicated \gls{emci} and \gls{emp} components \cite{eckerEMCIEMPDevelopmentsExperience2025}.
\end{itemize}

\subsection{The Measurement Challenge}
To accurately compare the latency of the \gls{mhfb} and \gls{mh}, the time difference between a \gls{can} message appearing on the bus and the corresponding data being output by the central controller must be measured. A naive approach using separate PC-based tools (e.g. a \gls{usb} and \gls{can} sniffer, as well as a \gls{uart}-to-\gls{usb} adapter) is insufficient. The non-deterministic latencies of the \gls{usb} bus and the PC's operating system would introduce significant and unpredictable errors, making a valid comparison impossible \cite{duEffectsPerceivedUSBdelay2016}.

\subsection{The Readout Hub Testbed Design}
To overcome the challenge of unpredictable timing errors introduced by conventional PC-based measurement tools, a dedicated "Readout Hub" testbed is proposed. The testbed is specifically designed to investigate the performance limitations of the Raspberry Pi-based prototype. It is predicted that the restrictions of the Pi's \gls{io} banks and the necessity of software-based switching between \gls{can} buses---required due to the lack of sufficient parallel hardware interfaces---will lead to data loss and increased latency as the message rate rises. Unlike the deterministic bus limit, this software bottleneck is non-trivial to calculate theoretically, necessitating experimental verification.

The design centers on a single microcontroller to ensure all timestamps are generated from a common, high-precision clock source, effectively eliminating relative timing errors. An STM32F767ZIT6 microcontroller is selected for this role due to its ARM Cortex-M7 core (running at up to 216 MHz) and its availability of 16 interrupt pins, which are essential for handling the SPI communication interface of the external CAN controllers simultaneously. Furthermore, it features high-resolution hardware timers, providing microsecond-level precision \cite{STM32F401RESTM32Dynamic}.

\begin{figure}[htbp]
\centering
\includegraphics[width=0.7\columnwidth]{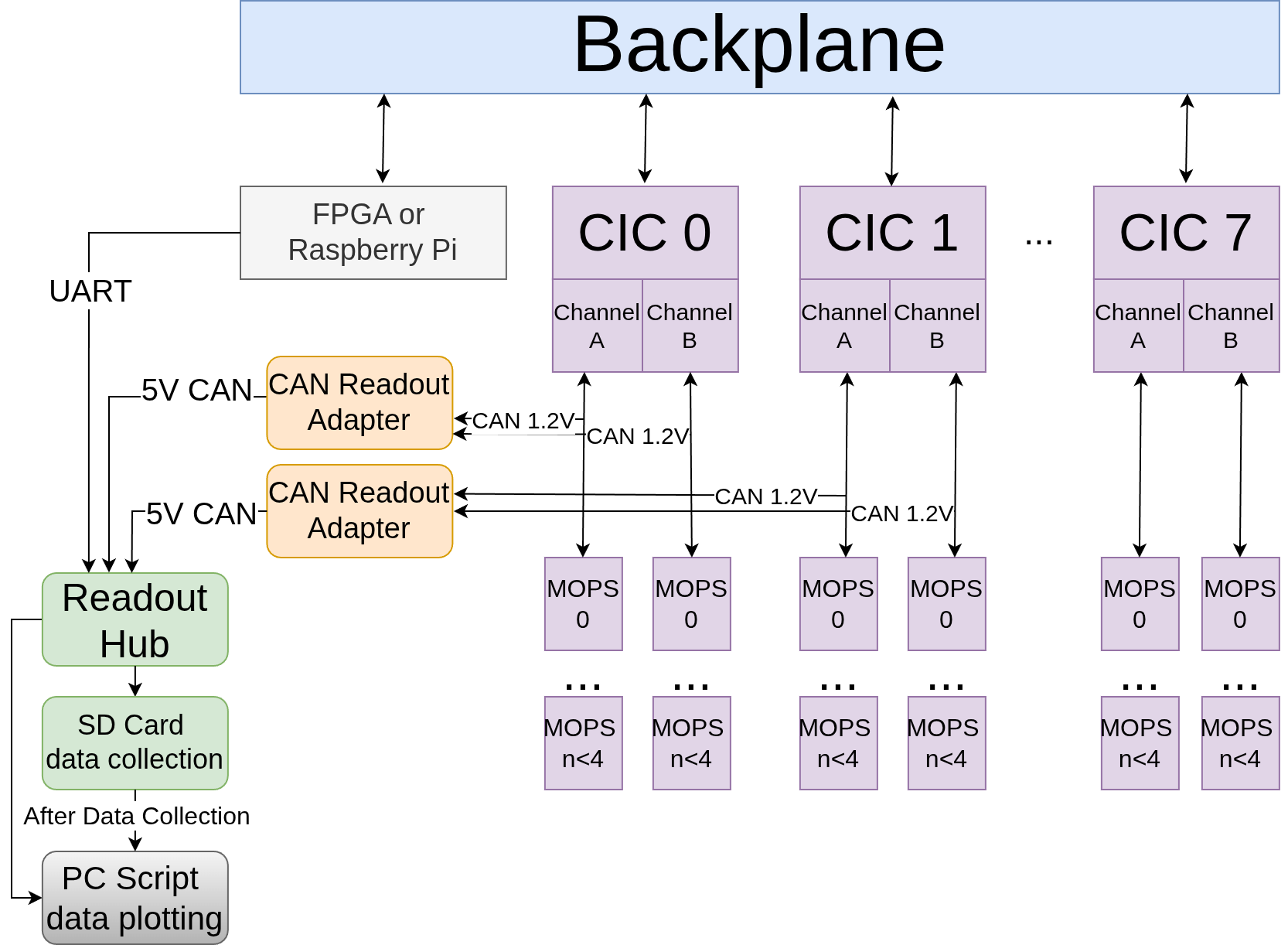}
\caption{The proposed general testbed architecture for latency and data integrity measurements.}
\label{fig:general_testbed_setup}
\end{figure}

\begin{figure}[htbp]
\centering
\includegraphics[width=0.8\columnwidth]{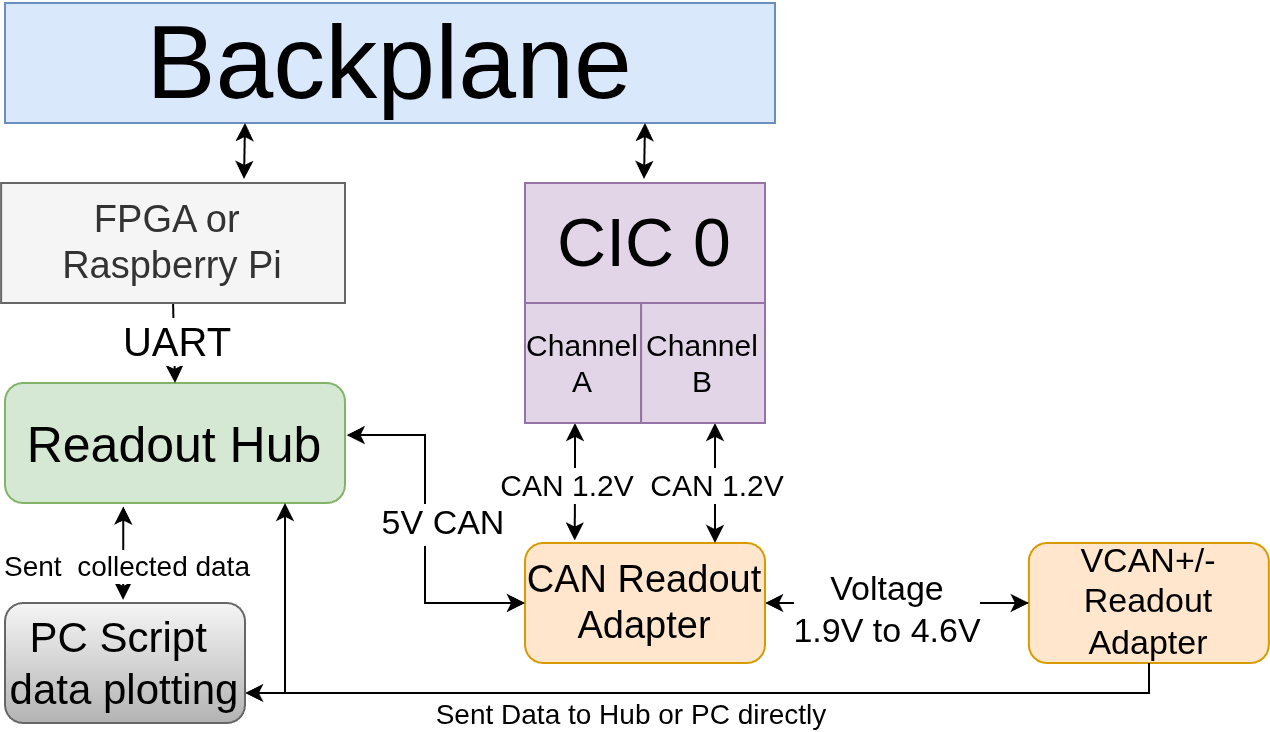}
\caption{The proposed testbed architecture for CIC functionality and isolation verification.}
\label{fig:cic_testbed_setup}
\end{figure}

The general testbed architecture is shown in Figure \ref{fig:general_testbed_setup}. A modified setup for \gls{cic} verification is shown in Figure \ref{fig:cic_testbed_setup}.

The key functional blocks of the testbed are:
\begin{itemize}
    \item \textbf{\gls{can} Readout Adapter}: This board serves as a non-intrusive tap into the 1.2V \gls{can} bus. Its primary component is a CAN-Bus Level-Shifter, which passively listens to the differential signals, converts them to 5V logic-level signals (RX), and passes them to the Readout Hub. The adapter also includes a transmit path (TX) for injecting messages onto the bus, a feature required specifically for the \gls{cic} functionality test. The board itself does not interpret the \gls{can} protocol, but it should have the capability to isolate its voltages and \gls{can} interface from other parallel CAN channels in the system to prevent measurement crosstalk.
    \item \textbf{Readout Hub (STM32 Core)}: The core of the measurement system. A custom \gls{pcb} houses the STM32 and external \gls{can} controllers (e.g. MCP2515). The STM32 firmware performs two primary tasks: (1) it receives decoded \gls{can} messages via SPI and records a high-precision timestamp ($t_1$); (2) it listens on the \gls{uart} debug port of the device under test for the processed data and records a second timestamp ($t_2$).
    \item \textbf{VCAN+/- Readout Adapter}: For the \gls{cic} functionality test the CAN readout Adapter can be used and should provide measuring probe points for VCAN+/-. To measure this an additional board is needed to measure the voltage differences, which then can be sent directly to the host PC or the Readout Hub.
\end{itemize}

\subsubsection{Data Logging and Statistical Analysis}
\begin{figure}[htbp]
    \centerline{\includegraphics[width=0.9\columnwidth]{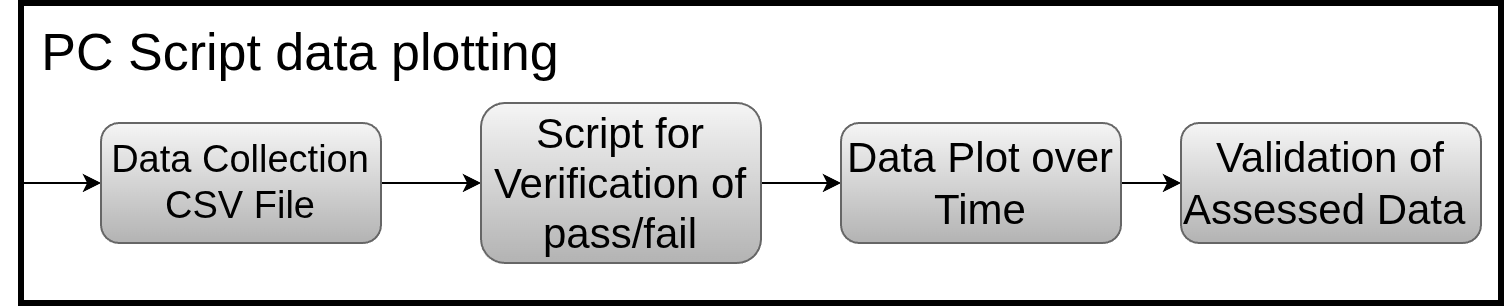}}
    \caption{The data logging and analysis pipeline, from raw data collection to final validation.}
    \label{fig:datalogging_flow}
\end{figure}

The collected data tuples ($t_1, t_2$, message payload) are streamed via \gls{usb} to a host PC for logging and analysis. This process, illustrated in Figure \ref{fig:datalogging_flow}, is a critical part of the methodology, turning raw timing data into verifiable performance metrics.

The pipeline consists of four main stages:
\begin{itemize}
    \item \textbf{Data Collection}: The Readout Hub saves all captured data (timestamps, message content) into a CSV file on a host PC, ensuring no data is lost and providing a complete record for post-processing.
    \item \textbf{Automated Verification}: As seen in Figure \ref{fig:datalogging_flow}, a Python script first parses the data to verify integrity (e.g. checking for packet loss) and performs an initial pass/fail check against the acceptance criteria defined in Section \ref{sec:evaluation}.
    \item \textbf{Statistical Analysis and Visualization}: The script then calculates the key performance metric, latency ($\Delta t = t_2 - t_1$), for every message. Latency is the most direct measure of the system's responsiveness. From this dataset, key statistics are derived, including the mean latency (average performance), standard deviation (jitter, a measure of predictability), and latency distribution over time. Visualizations, such as histograms and plots of latency over the test duration, are generated to help identify trends or anomalies.
    \item \textbf{Validation}: The generated statistics and plots provide the quantitative evidence needed to validate the system's performance against its requirements. This data-driven approach ensures an objective and repeatable assessment.
\end{itemize}

\subsection{Verification Techniques}
The verification methodology presented in this work is predicated on two core principles: measurement accuracy and comprehensive system assessment. The choice of specific techniques—a custom microcontroller-based testbed and a multi-stage evaluation plan—is a direct consequence of these principles. The selection of these communication-focused tests is crucial because reliable and timely data transfer is the fundamental prerequisite for any DCS functionality. Communication failures or performance degradation are often primary indicators of underlying hardware issues (e.g., power problems, faulty connections) or software bottlenecks. Therefore, rigorously verifying the communication channel provides essential insights into the overall system health and readiness, justifying this focus over other potential tests at this validation stage.

\subsubsection{Measurement Accuracy: The Case for a Dedicated Testbed}
A naive approach to measuring latency might involve separate, off-the-shelf, PC-based instruments, such as a commercial USB-to-CAN adapter and a USB-to-UART converter. However, this method introduces significant and, crucially, non-deterministic timing errors. As documented by Du et al. \cite{duEffectsPerceivedUSBdelay2016}, the inherent latencies of USB protocols and the scheduling uncertainties of non-real-time operating systems (e.g. Windows or standard Linux) can introduce jitter on the order of milliseconds. Such errors would be on the same scale as the expected latency of the system under test, rendering the measurements unreliable and making it impossible to confidently verify the performance criteria. The proposed "Readout Hub" built on a single STM32 microcontroller, circumvents this issue. By using a single, high-frequency clock source for all timestamps, this paper eliminates relative timing errors. This ensures that the measurements reflect the true performance of the MOPS-Hubs with a systematic uncertainty well below 10$\mu$s, as justified in Section \ref{sec:validity}. This level of precision is essential for reliably validating the millisecond-scale performance targets defined in Section \ref{sec:evaluation}.

\subsubsection{Comprehensive Assessment: The Three-Stage Evaluation Plan}
A single performance metric is insufficient to qualify a system for a critical application within the ATLAS DCS. A simple latency test might show good average performance but could miss critical flaws in scalability, long-term stability, or hardware isolation. Therefore, this paper adopted a structured, three-stage evaluation plan derived from established practices (see Chapter \ref{sec:related_work}) in high-reliability software and hardware testing \cite{naikSoftwareTestingQuality2008} to ensure a holistic assessment. Each test case is designed to yield specific conclusions about the system's capabilities:
\begin{enumerate}
    \item \textbf{Baseline Performance (Test Case 1):} This establishes a fundamental performance benchmark under ideal, low-load conditions. It provides the "best-case" latency (calculated from the mean of $\Delta t = t_2 - t_1$) and jitter (quantified by the standard deviation of $\Delta t$) figures. The deterministic nature of FPGAs, which is critical in physics experiments \cite{almeRadiationTolerantSRAMFPGABased2008, moopsHubFirmwareSpecificationV2}, is expected to yield very low jitter. Conclusion drawn: This test verifies the basic command-response functionality and quantifies the system's optimal communication speed and timing consistency.
    \item \textbf{Full Crate Stress Test (Test Case 2):} This test probes the system's robustness and scalability by subjecting it to the maximum specified data load, simulating the busiest operational conditions. The extended 8-hour duration is a form of "soak test," a standard industry practice for identifying long-term issues like memory leaks or thermal degradation \cite{SoakTestingWeb}. During this test, latency statistics are monitored over time to detect any performance degradation that might not appear in short-term tests. This is particularly important for the MHFB, where software-based bottlenecks under high throughput are anticipated. Conclusion drawn: This test validates the system's ability to handle the maximum expected data throughput without data loss or unacceptable latency increases, confirming its scalability and long-term stability.
    \item \textbf{CIC Isolation Test (Test Case 3):} This test moves beyond just data throughput, as in Test Case 1 and 2, to verify a critical hardware safety requirement: the electrical isolation between channels, mandated by the ITk grounding and shielding scheme \cite{ATLASITkGrounding}. Failure to verify this could lead to noise propagation and ground loops in the final installation. The technique of stressing one channel while monitoring another for crosstalk is a standard method for validating signal integrity and EMI resistance \cite{ISO1189812015}. Conclusion drawn: This test confirms that the CIC hardware meets the mandatory electrical isolation specifications, preventing potential noise issues in the final detector environment.
\end{enumerate}
This systematic approach ensures that the MH architecture is not only fast, but also scalable, stable, and compliant with the stringent electrical requirements of the ATLAS ITk environment.

\section{Measurement Validity and Uncertainty}
\label{sec:validity}
To ensure the credibility of the latency measurements, the systematic errors introduced by the testbed itself must be characterized and bounded. A brief calibration procedure will be performed to quantify these delays. This includes measuring the internal processing latency of the testbed from the moment a CAN frame is received by the MCP2515 controller to the SPI transaction servicing it, as well as the delay in capturing the first byte of a UART frame. The STM32's internal timer, driven by a high-frequency crystal oscillator, is expected to have negligible drift over the duration of a single test run (typically hours). Based on the microcontroller's clock speed and the SPI/UART baud rates, the combined systematic uncertainty of a single latency measurement ($\Delta t$) is expected to be well below 10$\mu$s. This level of precision is more than sufficient to resolve the anticipated millisecond-scale latencies of the systems under test. This expectation stems directly from the physical limitations of the 125 kbit/s CAN bus, where the transmission time for a single data frame is approximately 1.1~ms (as detailed in Section \ref{sec:evaluation}), establishing a baseline delay in the millisecond range before any processing overhead is considered \cite{STM32F401RESTM32Dynamic}.

\section{Evaluation Plan} \label{sec:evaluation}
The evaluation is structured into three main test scenarios derived from established hardware and software validation principles~\cite{naikSoftwareTestingQuality2008}. This structured approach is designed to systematically probe different aspects of system performance, from baseline characterization under ideal conditions to resilience under worst-case operational load and verification of specific hardware requirements~\cite{qameshQAQCMOPSHUB}. Each test is designed not only to quantify performance metrics but also to reveal potential failure modes specific to the different architectures (software-driven vs. FPGA-based). Each test will run for a specified duration to ensure the statistical significance of the results and to check for long-term stability issues, a common practice in qualifying high-reliability systems.

\subsection{Test Case 1: Baseline Performance} \label{subsec:case_1}
This test establishes the baseline round-trip latency for a single command-response transaction under minimal load.
\begin{itemize}
    \item \textbf{Setup}: One \gls{cic}, one channel, one \gls{mops} node.
    \item \textbf{Procedure}: A single \gls{sdo} read request will be initiated to query a register on the \gls{mops} node. The testbed will measure the round-trip latency from the transmission of the CAN request frame to the reception of the corresponding CAN response frame.
    \item \textbf{Repetitions and Duration}: The procedure is repeated 1,000 times to provide a statistically significant sample size for calculating the mean latency and its standard deviation (jitter). A duration of approximately 30 minutes is allocated for this test, which is sufficient to complete the repetitions and observe any immediate operational issues.
    \item \textbf{Acceptance Criteria}:
\begin{itemize}
    \item Packet loss must be zero.
    \item A maximum mean latency ($\overline{\Delta t}$) of 7~ms is defined as the acceptance criterion. Although the MH is a component of a ``slow control'' system where response times in the order of hundreds of milliseconds are functionally acceptable~\cite{poyDetectorControlSystem2008}, this stringent target is established as a critical performance benchmark. The theoretical minimum round-trip time ($t_{min}$) is dictated by the physical constraints of the 125~kbit/s bus~\cite{ahmadMonitoringPixelSystem2022, walsemannCANopenBasedPrototype2020a}.A worst-case 8-byte (64-bit) data frame utilizes approximately 135~bits on the bus, because the payload is expanded by 44~bits of protocol overhead for addressing and error checking~\cite{vossControllerAreaNetwork2008}. Additionally, the total size includes up to 27~bits for hardware-level bit stuffing and inter-frame spacing, which are essential for maintaining clock synchronization between nodes~\cite{vossControllerAreaNetwork2008}. This yields a one-way frame time ($t_{frame}$) of (\eqref{eq:y:1}:
    \begin{equation} \label{eq:y:1}
        t_{frame} = \frac{135~\text{bits}}{125,000~\text{bits/s}} \approx 1.1~\text{ms}.
    \end{equation}
    The minimum round-trip bus time is therefore $t_{min} = 2 \times t_{frame} \approx 2.2$~ms. Applying the required safety factor of 3 \cite{ahmadDescriptionManualUsea} results in 6.6~ms, which is rounded up to establish the 7~ms benchmark. This target allocates a processing budget of less than 4.8~ms for the entire electronic chain and confirms a highly responsive design with a significant performance safety margin ($7~\text{ms} / 2.2~\text{ms} \approx 3.18$), ensuring the system has ample headroom to handle processing overhead and system jitter without risk.
    \item The latency jitter ($\sigma_{\Delta t}$) for the MH must be tightly constrained. This serves as a benchmark to quantitatively validate the deterministic nature of the FPGA-based design, a crucial feature for reliable monitoring systems, as demonstrated in comparable high-energy physics electronics~\cite{almeRadiationTolerantSRAMFPGABased2008, moopsHubFirmwareSpecificationV2}.
    \end{itemize}
\end{itemize}
\subsection{Test Case 2: Full Crate Stress Test}  \label{subsec:case_2}
This test evaluates system scalability and performance under the maximum specified data load.
\begin{itemize}
    \item \textbf{Setup}:
        \begin{itemize}
            \item \textbf{\gls{mhfb}}: Maximum configuration of 4 \glspl{cic} (8 channels total), each with 1 \gls{mops} node \cite{lezkiMOPSHubBeginnersDocumentationa}.
            \item \textbf{\gls{mh}}: Full crate configuration of 8 \glspl{cic} (16 channels total), each with 4 \gls{mops} nodes (64 nodes total) \cite{qameshQAQCMOPSHUB}.
        \end{itemize}
        \item \textbf{Procedure}: All connected MOPS nodes will be configured to autonomously stream \gls{pdo} messages at a specified maximum rate. The MOPS ASIC's data rate is 125 kbit/s and the maximum message rate, as well as the round trip should not exceed the calculation made for \textit{Test Case 1}.
        Furthermore, the testbed will measure the one-way latency for each message.
        \item \textbf{Duration}: 8 hours per system. This extended duration is a standard reliability engineering practice known as a "soak test" designed to identify long-term stability issues such as thermal creep, memory leaks, or performance degradation under continuous, maximum load \cite{naikSoftwareTestingQuality2008, SoakTestingWeb}.
        \item \textbf{Duration}: 8 hours per system. This extended duration is a standard reliability engineering practice known as a "soak test" designed to identify long-term stability issues such as thermal creep, memory leaks, or performance degradation under continuous, maximum load \cite{naikSoftwareTestingQuality2008, SoakTestingWeb}. The primary objective of a "soak test" is to simulate real-world operational stress, observing the system's behavior over a prolonged period to uncover latent defects that might not manifest in shorter test cycles. Such tests are particularly effective at revealing resource-related bugs, where minor issues like un-freed memory accumulate over time, eventually leading to system failure.
        \item \textbf{Acceptance Criteria}:
        \begin{itemize}
            \item Packet loss must remain zero. This is a non-negotiable criterion, as any loss of monitoring data could mask a developing fault condition.
            \item The average one-way latency ($\overline{\Delta t}$) must remain below 3.3~ms (1.1~ms x safety factor of 3). This target is higher than the baseline 7~ms \eqref{eq:y:1} round-trip value because it accounts for the significantly increased processing load of aggregating and forwarding data from 64 nodes simultaneously. It verifies that the hub's architecture avoids becoming a data bottleneck even under worst-case "data storm" conditions.
            \item To ensure consistent performance, the 99.9th percentile of all latency measurements must not exceed 7~ms. This criterion formally defines the acceptable limit for outliers, permitting a statistically insignificant number of anomalous delays (fewer than 1 in 1000 packets) while guaranteeing that the system's response is highly predictable and reliable for the vast majority of operations.
            \item The extended 8-hour duration constitutes a  "soak test" a standard reliability engineering practice designed to identify long-term stability issues such as thermal creep, memory leaks, or performance degradation under continuous, maximum load~\cite{naikSoftwareTestingQuality2008, SoakTestingWeb}.
        \end{itemize}
\end{itemize}

\subsection{Test Case 3: CIC Functionality and Isolation Test}  \label{subsec:case_3}
This test verifies the electrical and logical isolation between the two channels of a single \gls{cic}, a critical requirement of the ITk grounding and shielding scheme \cite{ATLASITkGrounding}.
\begin{itemize}
    \item \textbf{Setup}: One \gls{cic} with its two channels (A and B) monitored by the testbed, as shown in Fig. \ref{fig:cic_testbed_setup}.
    \item \textbf{Procedure:} A continuous stream of CAN frames will be injected into Channel A at 200~Hz. This rate is intentionally chosen to be significantly higher than the maximum specified \gls{mops} data rate to aggressively stress the physical layer's driver, receiver, and termination components. This method of testing beyond nominal parameters is a standard approach to uncover weaknesses in signal integrity and susceptibility to electromagnetic interference (EMI)~\cite{ISO1189812015}. Simultaneously, Channel B will be monitored for any spurious data (crosstalk). The absence of signals on Channel B validates the electrical isolation, which is critical for preventing ground loops and noise propagation between different detector segments~\cite{ATLASITkGrounding}. This procedure aligns with the standard hardware validation checks for the CIC module~\cite{qameshQAQCMOPSHUB}. The roles of the channels are then reversed to ensure symmetrical performance.

    \item \textbf{Duration}: 30 minutes per \gls{cic} to confirm robust electrical isolation.
    \item \textbf{Acceptance Criteria}:
    \begin{itemize}
        \item Zero packets shall be detected on the inactive (monitored) channel.
        \item The common-mode voltage on both channels must remain within the electrical tolerances specified \cite{ahmadDescriptionManualUsea}.
    \end{itemize}

\end{itemize}

\section{Discussion and Expected Results}
\label{sec:discussion}

The evaluation is anticipated to yield a clear quantitative distinction between the \gls{mhfb} and \gls{mh} systems.

For the \gls{mhfb}, the Raspberry Pi is expected to be the primary performance bottleneck. The theoretical limitation of this architecture lies in its reliance on software-based context switching (multiplexing) to handle multiple CAN buses \cite{carvalho2019raspberry, duEffectsPerceivedUSBdelay2016}. Unlike a parallel hardware architecture, the Raspberry Pi must time-slice its processing resources to listen to and send data across different channels. It is hypothesized that during high-load scenarios, such as the full crate stress test, the overhead introduced by this multiplexing—combined with the non-deterministic scheduling of the Linux operating system—will lead to situations where information is lost because the processor is engaged with another channel when a new message arrives. While the \gls{mhfb} has been invaluable for initial development and validation, these tests are expected to confirm its unsuitability for the final production system due to these inherent latency and data integrity risks.

For the \gls{mh}, the \gls{fpga}-based architecture is expected to demonstrate superior performance due to its ability to process data from multiple channels in a truly parallel manner. In this architecture, the computational processing time is negligible; therefore, the system is not limited by the controller's performance, but rather by the physical transmission speed of the bus. The limiting factor is expected to be the theoretical minimum round-trip time of approximately 2.2 ms mandated by the 125 kbit/s CAN baud rate\cite{ahmadDescriptionManualUsea, vossControllerAreaNetwork2008}, rather than any internal processing bottleneck. Consequently, the latency is expected to be significantly lower and, crucially, deterministic with very low jitter \cite{davisSurveyHardRealtime2011}. In the full crate stress test, it is hypothesized that the \gls{mh} will handle the maximum data load from 64 \gls{mops} nodes with zero packet loss under a non-irradiation environment. The results should validate the custom \gls{fpga} platform as a robust, scalable, and reliable choice for the demanding environment of the \gls{itk} \gls{dcs}.

The successful completion of these tests will provide the necessary evidence to approve the \gls{mh} design for the production phase. The detailed performance data will form a crucial part of the final design review and will give confidence in the long-term stability and reliability of the \gls{mops} readout chain.

\section{Conclusion}
\label{sec:conclusion}
This paper has presented a comprehensive and rigorous methodology for the evaluation and validation of the final \gls{mh} readout system for the ATLAS \gls{itk} \gls{dcs}. The work has detailed the architecture of the \gls{mops} data path, highlighted the anticipated limitations of the initial prototype, and designed a dedicated measurement testbed using a real-time microcontroller to eliminate external sources of latency and error. The evaluation plan provides a clear, repeatable, and quantitatively-grounded procedure that includes baseline performance tests, high-load stress tests, and specific functional tests for key components. The successful execution of this methodology will provide the necessary data to quantitatively validate the performance, scalability, and reliability of the FPGA-based \gls{mh} system against the stringent requirements of the \gls{hl-lhc} environment. This verification is a critical step towards the production and final installation of the \gls{mops} system.

\section{Future Work}
\label{sec:future_work}
The work presented in this paper lays the foundation for the final validation stages of the \gls{mops} DCS-readout system. The immediate future work, which will form the basis of a Master's thesis, involves the physical development and implementation of the described testbed to execute the full evaluation plan. Following the successful validation of the \gls{mh} crate, the next phase will involve integrating it with the subsequent stage of the data acquisition chain: the \gls{irb} running an \gls{opcua} server. This integration is a critical step towards verifying the full data path as envisioned for the final system, bridging the gap between the custom front-end hardware and the wider CERN \gls{dcs} infrastructure \cite{dworakNewCERNControls2012}. The ultimate goal is to conduct a "full crate test" with the complete, production-like data chain, from the \gls{mops} \glspl{asic} to the \gls{opcua} server, thereby verifying the entire system end-to-end before deployment in the ATLAS experiment. The results of this complete test campaign will be presented in a forthcoming publication.

\acknowledgments

The authors would like to acknowledge the support of several institutions and individuals. Sincere thanks are extended to the University of Wuppertal for the collaboration and the provision of essential hardware. This research was supported by the laboratory facilities of the University of Applied Sciences Cologne (TH Köln). The productive cooperation with the ATLAS ITk MOPS development team is also gratefully acknowledged.

\bibliographystyle{JHEP}
\bibliography{ref}

\end{document}